\documentclass[pre,twocolumn,superscriptaddress]{revtex4}

\usepackage{graphicx}

\begin{document}

\setlength{\parskip}{0pt}
\setlength{\tabcolsep}{6pt}
\setlength{\arraycolsep}{2pt}

\title{Finding communities in linear time: a physics approach}
\author{Fang Wu}
\affiliation{Applied Physics Department, Stanford University,
Stanford, CA 94305}
\author{Bernardo A. Huberman}
\affiliation{HP Labs, Palo Alto, CA 94304}
\begin{abstract}
We present a method that allows for the discovery of
communities within graphs of arbitrary size in times that scale
linearly with their size. This method avoids edge cutting and is
based on notions of voltage drops across networks that are both
intuitive and easy to solve regardless of the complexity of the
graph involved. We additionally show how this algorithm allows for
the swift discovery of the community surrounding a given node
without having to extract all the communities out of a graph.
\end{abstract}
\maketitle

\section{Introduction}

The possibility of automatically discovering communities in large
network systems opens a promising set of new research areas in a
number of knowledge domains. From informal social networks that
can be discovered through their communication patterns \cite{huberman2}
to genetic clusters that lie hidden in the
biological literature \cite{huberman1} the unveiling
of community structures within these networks allows for the
investigation of information flow within organizations, the
discovery of causal effects in complex gene networks and the
dynamics of virus propagation in computer networks.

A central issue in the automatic discovery of communities is the
type of algorithms to be used with very large graphs, many of
which display a scale free structure. Not only are there problems
with the definition of communities per se, but also with the speed
with which these algorithms can uncover these communities.

By finding community structure within a network we mean that a
graph can be divided into groups so that edges appear within a
group much more often than across two groups. But this apparently
natural definition of community is problematic if a node connects
two clusters that have about the same number of edges. In this
case if becomes hard to tell to which cluster the node belongs.
Furthermore, large graphs often possess a hierarchical community
structure and hence the number of communities in a graph depends
on the level at which the graph is being partitioned.

Concerning the type of algorithms that have been used to discover
community structure, a recent one that has been used is based on
the idea of betweenness centrality, or betweenness, first proposed
by Freeman \cite{freeman}. The betweenness of an edge is defined
as the number of shortest paths that traverse it. This property
distinguishes inter-community edges, which link many vertices in
different communities and have high betweenness, from
intra-com\-mu\-ni\-ty edges, whose betweenness is low. The original
algorithm, developed by Girvan and Newman \cite{girvan_newman}, 
was also extended to
gene community discovery by Wilkinson and Huberman
\cite{huberman1,huberman2}, who partition a graph into discrete
communities of nodes using random sampling techniques. In these
cases, the time involved to discover the community structure of
the graph scales as $O(n^3)$.

More recently, Newman and Girvan \cite{newman_girvan} proposed a
different technique, which focuses on currents flowing on edges of
a network in order to discover communities.In this edge cutting
algorithm the time involved, is of order $O(n^4)$, with $n$ the
number of nodes in the graph. This is because it first calculates
a matrix inverse, which usually takes $O(n^3)$ time and then it
computes the voltage vector, $V$, for each possible source/sink
pair resistor networks. These polynomial scalings make these
algorithms hard to use when computing the community structure of
very large graphs.

In the computer science literature, there are a number of fast heuristics,
such as ``FM-Mincut''\cite{kernighan_lin,fiduccia} that can
cluster a graph in linear time. However, since their approach consists in
breaking up a graph by recursively cutting it so as to end up with
the desired number of partitions, they are inefficient when trying
to find out the community around a given node.

In this paper we present a different method that allows for the
discovery of communities within graphs of arbitrary size in times
that scale linearly with their size ($O(V+E)$). This method avoids
edge cutting and is based on notions of voltage drops across
networks that are both intuitive and easy to solve regardless of
the complexity of the graph involved. We additionally show how
this algorithm allows for the discovery of a community surrounding
a given node without having to extract all the communities out of
a graph.

In what follows we present the algorithm in the context of a very
simple problem, and then extend it to the general case. We then
apply it to problems that have been considered earlier using much
slower algorithms, such as membership in Karate clubs and the
discovery of conferences within US college football data. Finally
we exhibit the power of our method in the discovery of communities
around given nodes without having to compute the full community
structure of the graph, and we test it on email data collected
from HP laboratories. A final section discusses these results and
outlines possible applications.

\section{A graph as an electric circuit}

We start by exhibiting the workings of this algorithm in the
simplest problem, i.e, how to divide a graph into two communities.
We then extend our method to more general $n$-community graphs.
Consider a graph $G=(V,E)$. Suppose we already know that node $A$
and $B$ belong to different communities, which we call $G_1$ and
$G_2$ (we will talk later what if we do not have this information
beforehand). The idea is that we imagine each edge to be a
resistor with the same resistance, and we connect a battery
between $A$ and $B$ so that they have fixed voltages, say 1 and 0.
Having made these assumptions the graph can be viewed as an
electric circuit with current flowing through each edge
(resistor). By solving Kirchhoff equations we can obtain the
voltage value of each node, which of course should lie between 0
and 1. We claim that, from a node's voltage value we are able to
judge whether it belongs to $G_1$ or $G_2$. More specifically, we
can say a node belongs to $G_1$ if its voltage is greater than a
certain threshold, say 0.5, and it belongs to $G_2$ if its voltage
is less than that threshold.

\subsection{Why it works}

\begin{figure}
   \centering\resizebox{0.75\columnwidth}{!}{\includegraphics{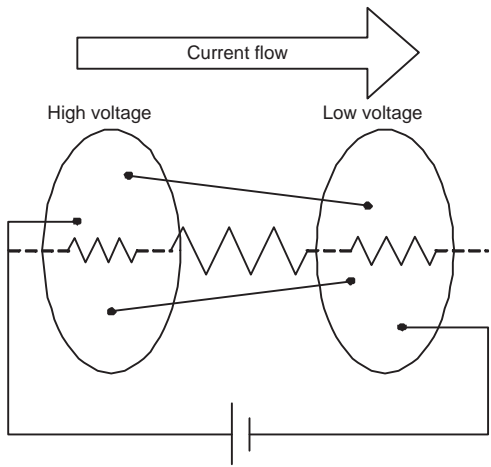}}
   \caption{Current flows from left to right, thereby building a
   voltage difference. Physically thinking, because nodes inside
   a community are densely connected, their voltages tend to be
   close. A big voltage gap happens about halfway between the two
   communities, where the edges are sparse and the local
   resistance is large.}\label{voltage_difference}
\end{figure}

\begin{figure}
   \centering\resizebox{0.75\columnwidth}{!}{\includegraphics{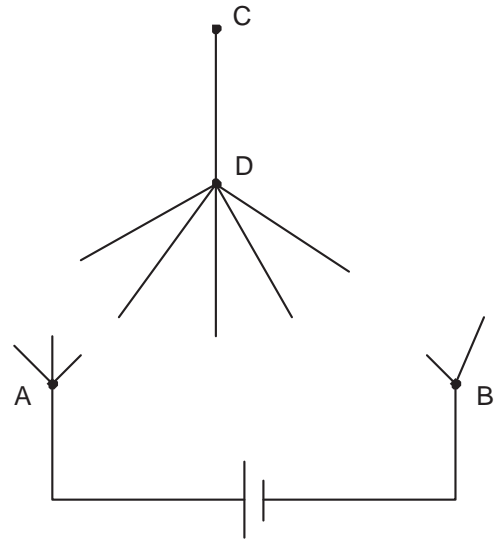}}
   \caption{A node with degree 1.}\label{degree1}
\end{figure}

First let us consider the simplest case that node $C$ has only one neighbor $D$,
so logically $C$ should belong to the same community as $D$ (Fig.~\ref{degree1}). 
Our idea indeed applies to this case. Because no current can flow through the 
edge $CD$, the two endpoints must have the same voltage, thus they belong to 
the same community.

Next we consider the case that node $C$ connects to two neighbors $D$ and $E$.
Because the edges $CD$ and $CE$ have the same resistance, we must have
$V_C=(V_D+V_E)/2$. Hence if $D$ and $E$ belong to the same community, i.e.,
$V_D$ and $V_E$ both lie above or below the threshold, then $V_C$ lying between
$V_D$ and $V_E$ should be above or below the threshold as well, therefore
belonging to the same community as $D$ and $E$, which makes sense. On the other
hand, if $D$ and $E$ belong to different communities, then it is comparatively
hard to tell which community $C$ belongs to ($V_C$ might be near the threshold),
but this is exactly where ambiguity arises - a node has connections with more
than one communities.

Last we consider the most general case: $C$ connects to $n$ neighbors $D_1,
\dots, D_n$. The Kirchhoff equations tell us the total current flowing into $C$
should sum up to zero, i.e.,
\begin{equation}
   \sum_{i=1}^n I_i = \sum_{i=1}^n \frac{V_{D_i}-V_C}R = 0,
\end{equation}
where $I_i$ is the current flowing from $D_i$ to $C$. Thus
\begin{equation}
   \label{neighboraverage}V_C = \frac 1n \sum_{i=1}^n V_{D_i}.
\end{equation}
That is, the voltage of a node is the average of its neighbors. If
the majority of $C$'s neighbors belongs to a community which has
voltage greater than the threshold, then $V_C$ tends to exceed the
threshold as well, hence our method tends to classify $C$ into
that community.

Our method can be easily extended to
weighted graphs. All we need to do is to set each edge's
conductivity proportional to its weight:
\begin{equation}
R_{ij} = w_{ij}^{-1}.
\end{equation}
The average appearing in Eq.~(\ref{neighboraverage}) becomes weighted
average accordingly.

\section{Kirchhoff equations in the general form}

Following Eq.~(\ref{neighboraverage}), the Kirchhoff equations of a $n$-node
circuit can be written as:
\begin{eqnarray}
\label{battery1}V_1 &=& 1,\\
\label{battery2}V_2 &=& 0,\\
\label{kirchhoff1}V_i &=& \frac 1{k_i} \sum_{(i,j)\in E} V_j =
\frac 1{k_i} \sum_{j\in G} V_j \,a_{ij} \quad\mbox{for }i=3,\dots,n,
\end{eqnarray}
where $k_i$ is the degree of node $i$ and $a_{ij}$ is the
adjacency matrix of the graph. Without loss of generality, we have
labelled the nodes in such a way that the battery is attached to
node 1 and 2, which we call \emph{poles}, accordingly
Eq.~(\ref{battery1}) and (\ref{battery2}). Eq.~(\ref{kirchhoff1})
is a set of linear equations of $n-2$ variables $V_3, \dots,
V_{n}$ that can be put into a more symmetrical form:
\begin{equation}
V_i=\frac 1{k_i} \sum_{j=3}^n V_j \,a_{ij} + \frac 1{k_i} a_{i1}
\quad\mbox{for }i=3,\dots,n.
\end{equation}
Define
\begin{equation}
V=\left(
   \begin{array}{c}
   V_3\\
   \vdots\\
   V_n
   \end{array}
\right), \quad B=\left(
   \begin{array}{ccc}
   \displaystyle\frac{a_{33}}{k_3} & \dots & \displaystyle\frac{a_{3n}}{k_3}\\
   \vdots & & \vdots\\
   \displaystyle\frac{a_{n3}}{k_n} & \dots & \displaystyle\frac{a_{nn}}{k_n}
   \end{array}
\right), \quad C=\left(
   \begin{array}{c}
   \displaystyle\frac{a_{31}}{k_3}\\
   \vdots\\
   \displaystyle\frac{a_{n1}}{k_n}
   \end{array}
\right),
\end{equation}
then the Kirchhoff equations can be further put into a matrix form
\begin{equation}
V=BV+C,
\end{equation}
which has the unique solution
\begin{equation}
\label{solution}V=(I-B)^{-1}C.
\end{equation}

In general it takes $O(n^3)$ time to solve a set of equations like
Eq.~(\ref{solution}). However, we can actually cut the
time down to $O(V+E)$, as described in the next section.

Before closing we point out that if we define
\begin{equation}
L=\left(
   \begin{array}{cccc}
   k_3 & -a_{34} & \cdots & -a_{3n}\\
   -a_{43} & k_4 & \cdots & -a_{4n}\\
   \cdots & & & \cdots\\
   -a_{n3} & -a_{n4} & \cdots & k_n
   \end{array}
\right), \quad D=\left(
   \begin{array}{c}
   a_{31}\\
   \vdots\\
   a_{n1}
   \end{array}
\right),
\end{equation}
then the Kirchhoff equations can also be written as
\begin{equation}
LV=D,
\end{equation}
which has the unique solution
\begin{equation}
V=L^{-1}D.
\end{equation}
Interestingly enough, $L$ is the Laplacian matrix of the subgraph
of $G$ containing nodes $3, \dots, n$. The well-known spectral
partitioning method partitions the graph based on the eigenvector
of the second smallest eigenvalue of $G$'s Laplacian matrix
\cite{psl,fiedler,fiedler2}. We point out however that our method
does not compute the eigenvectors of $G$.

\section{Solving Kirchhoff equations in linear time}

We first set $V_1=1, V_2=\cdots=V_n=0$ in $O(V)$ time. Starting
from node 3, we consecutively update a node's voltage to the
average voltage of its neighbors, according to Eq.~(\ref{neighboraverage}). 
The updating process ends when we
get to the last node $n$. We call this a round. Because any node
$i$ has $k_i$ neighbors, one has to spend an amount of $O(k_i)$
time calculating its neighbor average, thus the total time spent
in one round is $O(\sum_{i=3}^n k_i)=O(E)$. After repeating the
updating process for a finite number of rounds, one reaches an
approximate solution within a certain precision, which \emph{does
not depend on the graph size $n$ but only depends on the number of
iteration rounds}. In other words, to obtain a certain precision,
say 0.01, one only needs to repeat, say, 100 rounds, no matter how
large the graph is, so the total running time is always $O(V+E)$.

To show conceptually the fast convergence of the algorithm, we
expand Eq.~(\ref{solution}) into a series:
\begin{equation}
V=\sum_{m=0}^\infty B^m C.
\end{equation}
Now if we define
\begin{equation}
f(V)=BV+C
\end{equation}
then
\begin{equation}
f^{(r)}(V)=\sum_{m=0}^{r-1}B^mC + B^rC.
\end{equation}
As $r\to 0$ the remainder $\to 0$, so we see the iteration
algorithm amounts to a simple cutoff of the power series. The
convergence speed is determined by the matrix norm $||B||$ which
is usually insensitive to $\dim(B)=O(V)$.

\section{A two-community example: Zachary's karate club}

\begin{figure}
   \centering\resizebox{0.8\columnwidth}{!}{\includegraphics[scale=.5]{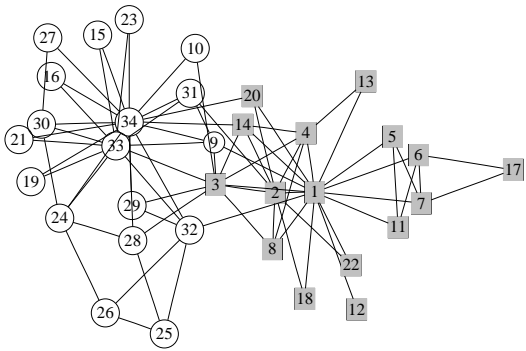}}
   \caption{Zachary's karate club. This figure is from Newman and 
   Girvan \cite{newman_girvan}.}\label{zachnet}
\end{figure}

\begin{figure*}
   \centering\includegraphics[scale=.75]{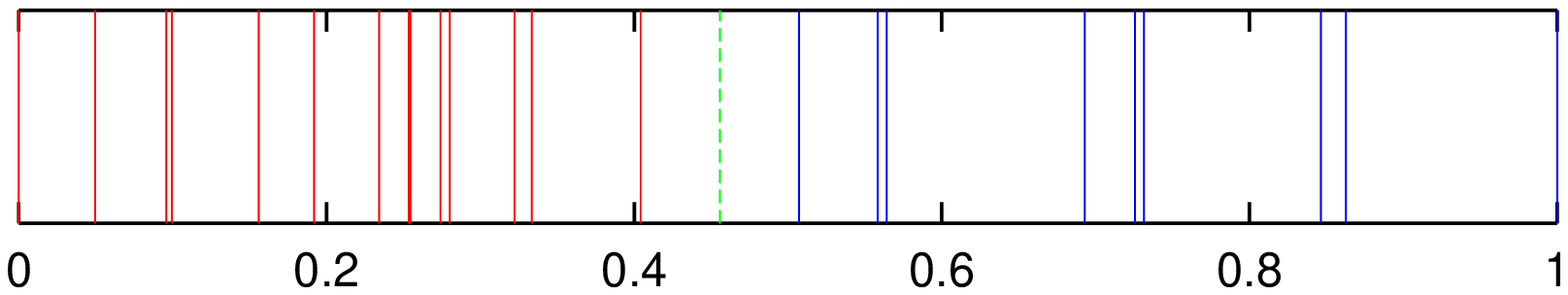}
   \centering\includegraphics[scale=.75]{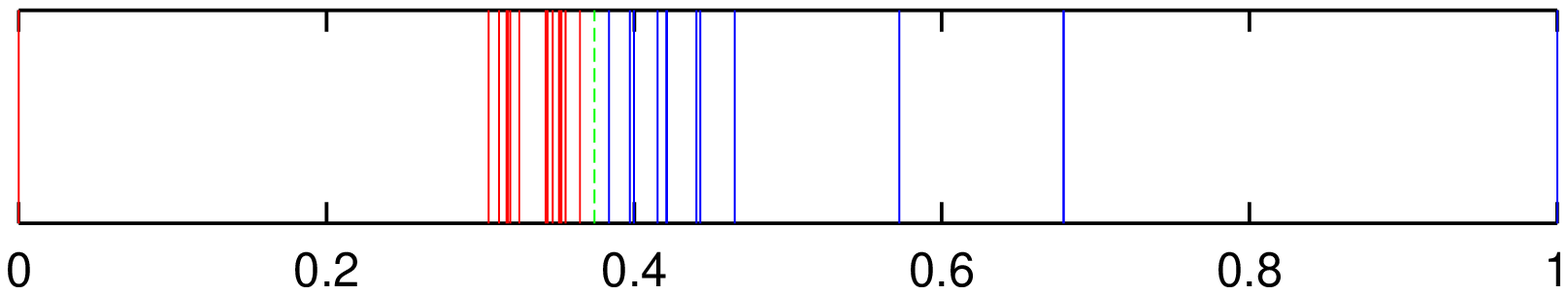}
   \centering\includegraphics[scale=.75]{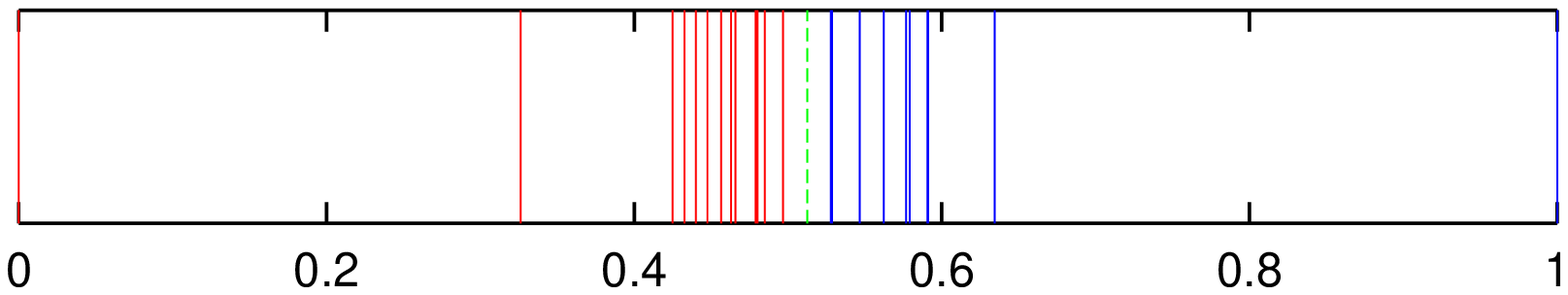}
   \centering\includegraphics[scale=.75]{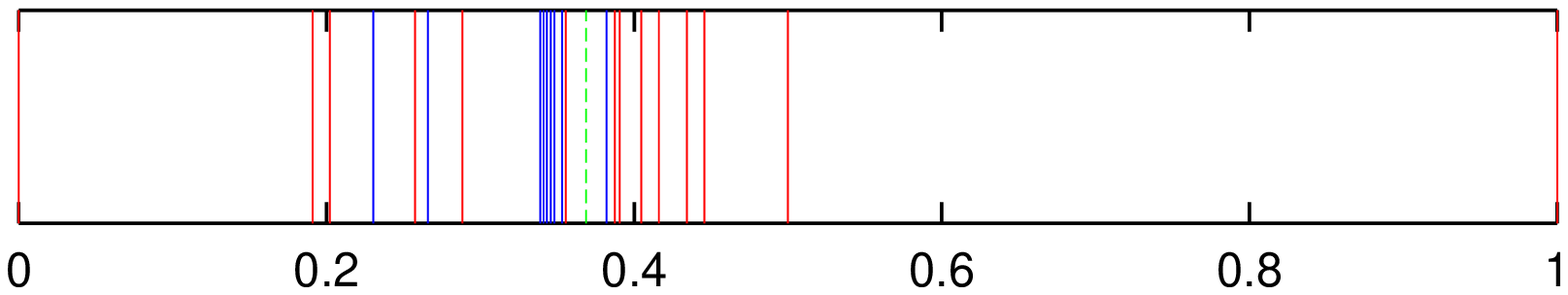}
   \caption{Voltage spectrum for the two community example.
   In the four panels the battery
    is hooked up to nodes (a) 1 and 34; (b) 16 and 17; (c) 12 and 26;
   and (d) 32 and 33. The algorithm runs 100 iteration rounds to reach the
precision $< 0.01$. Red lines and blue lines distinguish different
   communities (based on real data). Each graph is cut into two
   halves at the biggest gap near the middle (tolerance $=0.2$), which we marked
out with a green dashed line. As can be seen, the algorithm correctly
   recognizes the two communities when the two poles are
   in different communities ((a)--(c)), and fails when they belong
   to the same community.}\label{club}
\end{figure*}

We tested our algorithm against the friendship network data from
Zachary's karate club study \cite{zachary}. The
graph includes two communites of roughly equal size (Fig.~\ref{zachnet}). 
The results of our linear time algorithm are shown in Fig.~\ref{club}.

In the figures, a node is represented as a vertical line at the
abscissa equal to its voltage, and is painted either red if it
belongs to the first community, or blue if it belongs to the
second, based on Zachary's real data. If our algorithm works, the
red lines and the blue lines should separate at the two ends. This
is indeed the case for the first three examples, when the external
voltage is added between a pair of nodes lying in different
communities. We also show in the last panel how the algorithm
fails when the poles lie in the same community.

After obtaining the complete voltage spectrum two critical
questions remain to be answered:
\begin{itemize}
   \item How to pick the two poles so that they lie in different communities?
   \item What threshold should be used to separate the two communities?
\end{itemize}

The first question is hard because we do not have any prior
information about the graph and the problem has to be solved in
linear time. We first describe a heuristic that works although
inconsistently, and then present a better statistical method in
the next section.

Because nodes are densely connected inside a community, the
average distance between two nodes chosen from one community is
generally shorter than the average distance between two nodes
chosen from different communities. Thus, there is a high
probability that two far apart nodes sit in different communities,
qualifying for the poles.

To find a far apart pair of nodes one can use the following
linear-time method. First randomly pick a node, then find the node
farthest from it, using a simple breadth-first search which takes
time $O(V+E)$. If more than one node qualifies, pick any of them.
Next, use another breadth-first search to find the node farthest
from the second node, and so on. After a few steps this procedure
would identify a pair of nodes very far away.

The diameter of the graph is defined by the largest distance of
all pairs. The graph of the karate club has diameter 5. All pairs
of nodes with this distance apart indeed belong to different
communities. One example $(16,17)$ is shown in Fig.~\ref{club}(b).

The second question, i.e. what threshold to use in order to
separate the two communities, is easier to answer. Because edges
are sparser between two communities, the local resistivity should
be large compared to the local resistivity within the two
communities. Thus the voltage drops primarily at the junction (see
Fig.~\ref{voltage_difference}) between communities. This suggests
placing the threshold at the largest voltage gap near the middle.
Note that the global largest gap often appears at the two ends of
the voltage spectrum (see e.g.~Fig.~\ref{club}(b) and (c)), but it
does not make sense to cut there at all, which would divide the
graph into two extremely asymmetrical communities, one of which
has only one or two nodes. Of course this is not what we want.

To be more definitive, we now define rigorously the term ``near
the middle". We distinguish two cases:

\noindent 1.~Dividing the graph into exactly two equal-sized
communities.

We simply cut at the right middle gap. The median-selection problem can be done
in $O(V)$ time by a good selection algorithm \cite{cormen}.

\noindent 2.~Finding communities of roughly the same size, which
for the karate club example implies $\approx 34/2=17$ nodes each.

We define a \emph{tolerance} to describe the range of allowed
community sizes. A tolerance 0.2 means we only search for
communities of the size $17\pm 20\%$, which is $(14,21)$. First we
sort the voltage values. Then we find the the largest gap among
the middle $21-14=7$ gaps and cut there. Note that the sort can be
done in $O(V)$ time by using a standard linear time sort,
e.g.~counting sort \cite{cormen}, which applies to our problem
since the voltage can only take a finite number of values (101
choices for precision 0.01). The green dashed lines in
Fig.~\ref{club} were found this way.

\begin{figure}
   \centering\resizebox{0.75\columnwidth}{!}{\includegraphics{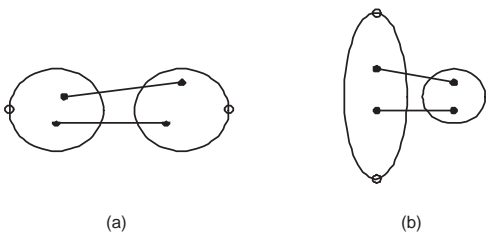}}
   \caption{(a) The largest distance happens across two
communites. (b) The largest distance can happen inside a community
sometimes.}\label{largest_distance}
\end{figure}

We emphasize that this method does not always work, as illustrated
in Fig.~\ref{largest_distance}.

\section{Choosing poles randomly}

A statistical method can be used to avoid the ``poles problem''
instead of solving it. The idea is to randomly pick two poles,
apply the algorithm to divide the graph into two communities, and
repeat it for many times (the total time is still $O(V+E)$). About
one half of the results would give correct results, for the poles
would happen to lie in different communities, while the other half
would give incorrect results. If we now improve our pole-picking
method by only choosing two nodes that are not neighbors (i.e.,
there is no edge between them), then the probability that our
randomly chosen poles lie in different communities becomes higher
than a half, suggesting the majority of the results is correct.
Thus we should be able to use a majority vote to determine the
communities.

\begin{figure}
   \centering\resizebox{0.75\columnwidth}{!}{\includegraphics{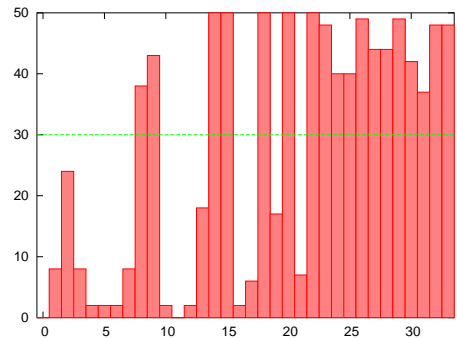}}
   \caption{The number of times a node appears in the same
group as node 16. There are altogether 50 groups containing node 16.}\label{majority_vote}
\end{figure}

We tested our method against the karate club data. Each time we
randomly picked two nodes whose distance $\ge 2$, and then ran the
algorithm to find two communities. We repeated the process 50
times to obtain 100 groups altogether, among which 50 groups
contained node 16 (16 has no special meaning - we arbitrarily
chose it). We counted, for each node, how many times it appeared
in the same group as node 16, the maximal possible value being 50
and the minimal value 0. The result is shown as a bar graph in
Fig.~\ref{majority_vote}. Comparing the graph with the real data
we see that those nodes in node 16's community indeed all have
high votings (above the green horizontal line in
Fig.~\ref{majority_vote}).

\section{Graphs with more than one communities}

We now extend our method to $n$-community graphs.  We test our algorithm against
the US college football data studied by Girvan and Newman \cite{girvan_newman}. A total
of 115 teams are divided into 13 ``conferences'' containing around 8 to 12 teams
each. Our task is to find all those conferences (communities).

As we proceeded in the karate club case, we first randomly pick
two poles whose distance $\ge 2$, then apply our algorithm to get
the voltage spectrum. (Note that the probability that two poles
belong to the same community decreases as the number of commuties
increases, roughly in the manner $1/m$, where $m$ is the number of
communities.) We set the tolerance to be 0.5, which means that we
only search for communties whose size is in the range $(115/13)\pm
50\%$, or between 4 and 13, roughly.

\begin{figure*}
   \centering\includegraphics{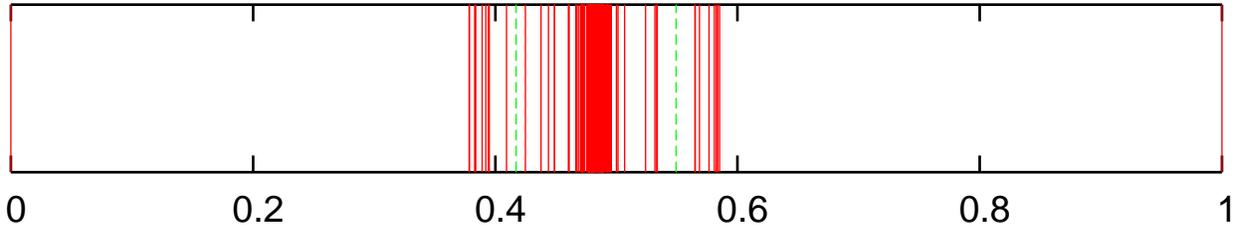}
   \caption{The voltage spectrum when the battery is
hooked up to node 51 (Washington) and node 88 (Tulsa). Two groups are identified
at the ends by green dashed lines.}\label{football_spectrum}
\end{figure*}

To be more precise, we sort all 115 voltage values in an increasing order and
label them as $0=V_1\le V_2\le\dots\le V_{115}=1$. We then measure the gaps
$V_6- V_5, V_7-V_6, \dots, V_{14}-V_{13}$ one by one to pick out the largest
one, say $V_9-V_8$, which indicates a group of nodes having voltages $V_1,
\dots, V_8$. Similarly, we obtain a group of nodes at the $V_{115}$ end. The two
groups thus found are both candidates for the 13 communities we are looking for.
An example is shown in Fig.~\ref{football_spectrum}.

\begin{figure*}
   \centering\includegraphics{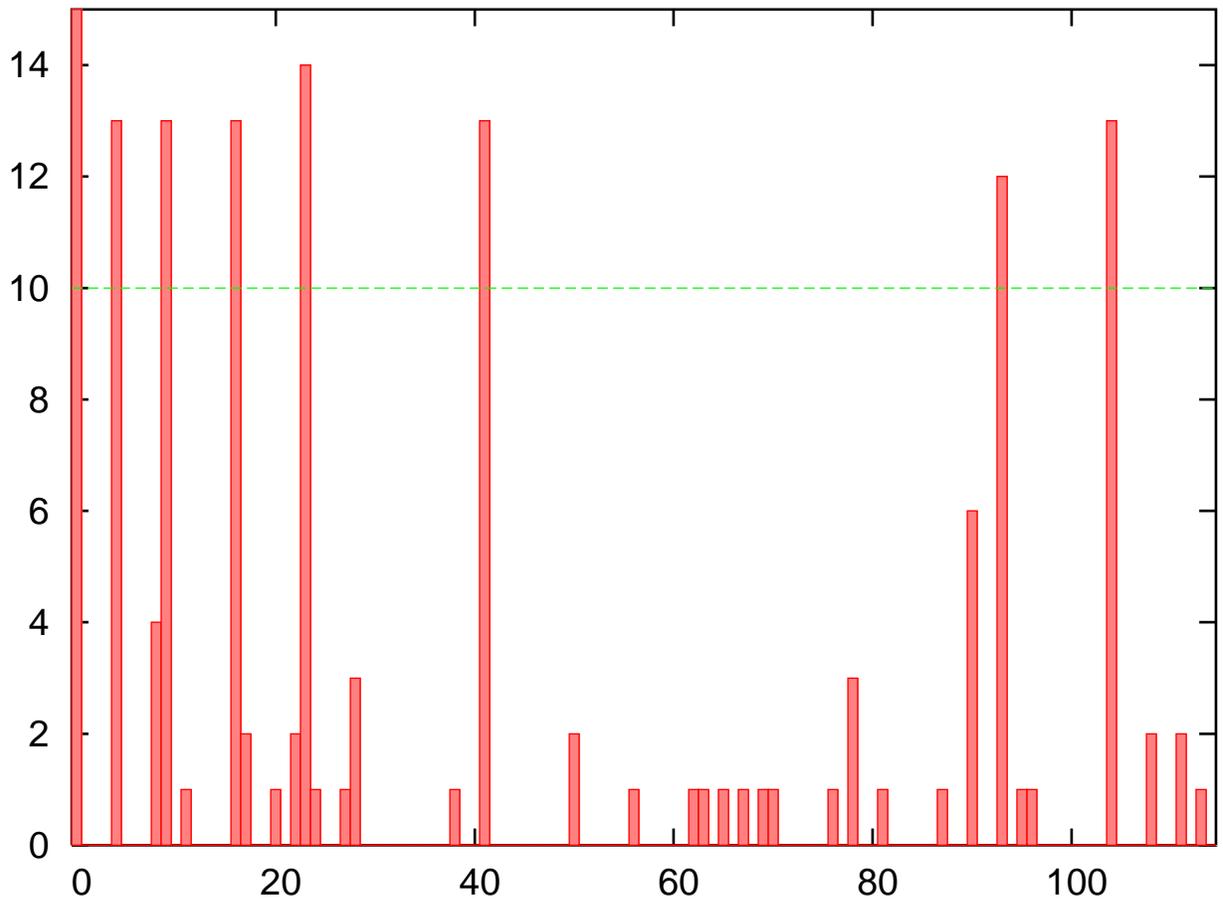}
   \caption{The number of times a node appears in the same
group as node 0 (Brigham Young). There are altogether 15 groups containing node
0. In the figure, eight nodes lie above the green threshold, namely node 0
(Brigham Young), 4 (New Mexico), 9 (San Diego State), 16 (Wyoming), 23 (Utah),
41 (Colorado State), 93 (Air Force), and 104 (Nevada Las Vegas). They are
exactly the members of the Mountain West conference.}\label{football_vote}
\end{figure*}

We repeated the process 50 times to collect 100 candidates. We
then found out all the groups containing a specific node to apply
a majority vote, just like what we did before to 2-community
graphs. The specific node can be chosen rather freely, but to use
most information, we chose the one that appears most frequently in
the 100 groups (frequency test takes $O(V)$ time). An example of
such a majority vote is shown in Fig.~\ref{football_vote}. After
we found the first community this way, we again picked a node in
the rest of the graph which appears most frequently, and applied a
majority vote to all groups containing that node in order to find
the second community. Repeating this procedure 13 times, we were
able to find out all 13 communities.

\section{Finding the community of a given node}

We can further save time if we are only required to find the
community of a given node instead of all communities. To this end,
instead of randomly picking two nodes at a time, we fix the given
node as one pole, and choose the second pole to be another random
node that is at least a distance 2 away from the first one. The
rest steps (setting the tolerance, calculating voltages, cutting
through the biggest gap, etc.) remain the same. By doing so each
round we are guaranteed to acquire a group containing the given
node, so we can further reduce the total number of rounds from 50
to, say, 20, which gives us 20 candidates, sufficient for the
majority vote.

We also tested our method against the HP labs email data, which
was collected from a roughly power-law network consisting of 396
nodes. We joined two nodes with an edge if the they exchanged more
than 30 emails a month. As an example, we tried to find out the
closest colleagues of the node ``Jaap''. Our results show a total
number of 20 nodes to lie above the threshold. Comparing this
result with the communities extracted from the email data, we
discovered that all these nodes belong to the same laboratory as
does the node Jaap, as was indeed the case.

\medskip\noindent\underline{Remark:} Distance information is not sufficient
to detect the community of a given node. One cannot simply pick out the nodes
within a radius $d$ from the given node and say they form a community, because
\begin{enumerate}
\item Two nodes separated by a short distance need not to be in the same
community. In our last example, 57 nodes have distance $\le 2$ from Jaap, among
which only 27 belongs to Jaap's lab.
\item For a small-world network, even the number of second neighbors or
third neighbors can be very large. In our last example Jaap has 157 neighbors
within a distance 3, which is already about 40\% of the total size.
\item Two nodes with a large distance apart can still be in the same
community. For example, ``JShan'' is among one of the 20 nodes found by our
algorithm but has a distance 3 away from Jaap, which is quite large.
\end{enumerate}

\section{Other interpretations of voltage}

In our 2-community example the voltage is regarded as an index
indicating which community a node belongs to. Its absolute value
has no special meaning, for we can freely change its range from
$[0,1]$ to any other range.

Despite its clear physical meaning, we can reinterpret the voltage
as a weight function measuring to what extent the node belongs to
a community. For example, if we set the voltage range to $[-1,1]$,
 we can then say a node ``strongly'' belongs to the $-1$
community if its voltage is $- 0.9$, or a node ``weakly'' belongs
to the 1 community if its voltage is $0.2$, etc.

This second interpretation of voltage inspires us to try other
possible choices of weight functions. The voltage, being a scalar,
can only separate two communities because the real line only has
two directions. If we generalize however our weight function to a
\emph{vector}, we can then achieve extra dimensions to separate
more communities.

\begin{figure}
   \centering\resizebox{0.75\columnwidth}{!}{\includegraphics{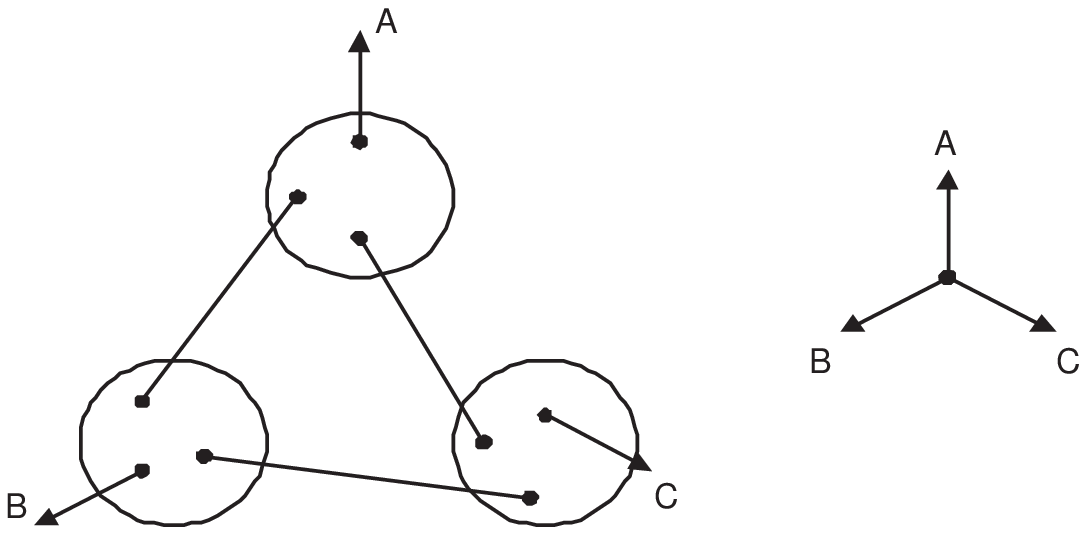}}
   \caption{A graph made of three communities. The three
sources lie in different communities. The angle between any two of
the weight vectors $\mathbf A$, $\mathbf B$ and $\mathbf C$ is 120
degree.}\label{vec_weight}
\end{figure}

For example, consider the 3-community graph in
Fig.\ \ref{vec_weight}. Suppose we have already found three poles
dispersed in three communities. We assign each pole a unit-length
vector weight in such a way that the angle between any two of the
them is exactly 120 degrees, shown in Fig.~\ref{vec_weight} as
$\mathbf A$, $\mathbf B$ and $\mathbf C$. Those vectors have the
nice properties $\mathbf A+\mathbf B=-\mathbf C$, $\mathbf
A+\mathbf C=-\mathbf B$ and $\mathbf B+\mathbf C=-\mathbf A$.
Thus, if a node is strongly connected to, say, communities
$\mathbf A$ and $\mathbf B$ but not to $\mathbf C$, then there is
a strong signal to separate the node from community $\mathbf C$
(because $\mathbf A+\mathbf B=-\mathbf C$). Also, if a node
connects to all three communities, we see that the relation
$\mathbf A+\mathbf B+\mathbf C=0$ indeed reflects the obscurity of
the node's belonging.

After we have fixed the vector weights of the three poles, we can
continue with our method to solve Kirchhoff equations. We only
need to replace the sums in Eq.~(\ref{kirchhoff1}) by vector sums.
Once we solve out the vector weights of all nodes, we can tell a
node belongs which community according to its pointing direction
in the 2-dimensional plane. For example, if a node's vector weight
is pointing basically upward then we can say it belongs to
community $\mathbf A$. Hence vector weights allow us to separate
three communities at a time.

While one might wish to further extend the method to higher
dimensional spaces to separate more communities at a time, we
point out that we have not yet succeeded in finding a symmetrical
set of vectors in three or higher dimensional spaces.

There is one more interesting probabilistic
interpretation of voltage \cite{rabinovich}: When a unit voltage
is applied between $a$ and $z$, making $V_a=1$ and $V_z=0$, the
voltage $V_x$ at any point $x\ne a,z$ represents the probability
that a walker starting from $x$ will return to $a$ before reaching
$z$. There is also a probabilistic interpretation of current.

\section{Discussion}

In this paper we presented a method that allows for the discovery
of communities within graph of arbitrary size in times that scale
linearly with its size. The method avoids edge cutting altogether
and is based on notions of voltage drops across networks that are
both intuitive and easy to solve regardless of the complexity of
the graph involved. Additionally, this method allows for the
discovery of a community surrounding a given node without having
to extract all the communities out of a graph.

We then tested the algorithm by applying it to several problems
such as membership in karate clubs and the discovery of
conferences within US college football data. We also show how it
can be used to discover of communities around given nodes by
working with a graph of email data collected from HP laboratories.

The reason behind the speed of this method lies in its focus on
communities themselves and not on their hierarchical structures.
In contrast, Newman's betweenness method \cite{girvan_newman}
detects not only the communities but also the complete hierarchy
tree using much longer times. While our method lacks the ability
to find the hierarchy tree, it also saves a lot of time since it
does not need to find out all the big communities before looking
for the small ones. In fact, it can identify the community of any
given node, without knowing the full structure of the graph or the
composition of other communities.

\begin{figure}
   \centering\resizebox{0.75\columnwidth}{!}{\includegraphics{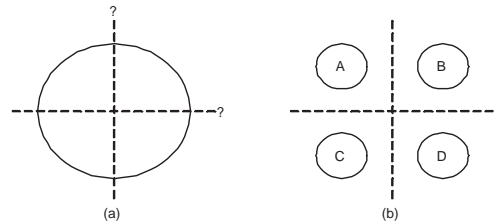}}
   \caption{Graph not divisible. (a) A graph that is
densely connected everywhere; (b) A graph made of four communities that are
about the same size (inter-community edges not shown).}\label{not_divisible}
\end{figure}

A possible defect of our method is that we have to specify the
number of communities we wish to divide the graph into, a piece of
information which one does not often have beforehand. A natural
solution would be to first divide the graph into two big
communities and then break them down into smaller ones by
recursively applying the method described before. Unfortunately,
the statistical method of attaching the battery to random sites
over the graph works poorly when the graph is not ``divisible''
enough, and this will happen whenever the graph itself is a big
community, and thus not divisible, or  when the graph can be
divided into two parts in many ways (``too divisible''), each
having about the same contribution to the majority vote
(Fig.~\ref{not_divisible}).

In order to explain why our statistical method works poorly in the
second case, consider the graph shown in
Fig.~\ref{not_divisible}(b), which is composed of four
communities, $A$, $B$, $C$ and $D$. Suppose $AB$, $AC$, $BD$ and
$CD$ are loosely connected by some inter community edges but not
$AD$ and $BC$. If we happen to choose two poles separately in $A$
and $B$, then our algorithm would tend to divide the graph into
two parts: $AC$ and $BD$. However, we have a roughly equal chance
to choose two poles in $A$ and $C$, which would imply the division
$AB$ and $CD$. Thus our statistical method becomes puzzled as to
where to cut.

We emphasize that, the reason our statistical method would fail
sometimes is due to the ambiguity of the graph itself. In our
previous example, any algorithm would and should hesitate whether
to cut the graph into $AB/CD$ or $AC/BD$. A good algorithm should
be able to yield at least one reasonable result. In fact, if we
are just interested in finding one solution, no matter which, we
could always apply the quick-and-dirty method by choosing two
poles far away. This would lead to a reasonable solution. In this
sense, our method might better be taken as a graph partitioning
method rather than a community detecting method.

In closing we point out a number of possible extensions of our
method that could make it even more effective when dealing with
complex graphs. The first one is a better statistical method that
still works well when the graph is ``too divisible''. Second, one
could also search for better weight functions and a better
definition of average other than the one in
Eq.~(\ref{kirchhoff1}). Third, there is information in the
complete voltage spectrum that has not yet been fully exploited.
For example, nodes belonging to the same community usually
concentrate closely in the spectrum, and yet the voltages between
the two green lines in Fig.~\ref{football_spectrum} were simply
discarded. Finally, one could use the result of a majority vote 
to evaluate the correctness of the partition.

In spite of lack of these extensions we believe that the algorithm
we have presented is fast and useful when trying to find
communities within large graphs or around a single node.

\bigskip We thank Zhao Wu and Li Zhang for some useful discussions.

\end{document}